# Nanoscale imaging reveals critical plating and stripping mechanisms in anode-free lithium and sodium solid-state batteries

J. Díaz-Sánchez,[a,b] P. Hernández-Martín,[a,b] N. Kwiatek-Maroszek,[c] H.R. Bratlie,[d] R. Anton,[e] A. Lowack,[e] A. Galindo,[f] K. Kataoka,[g] E. Vasco,[f] K. Nikolowski,[e] D. Rettenwander,[d] E.G. Michel,[a,b,h] M.A. Niño,[c] M. Foerster,[c] C. Polop[a,b,h*]

[a] *Condensed Matter Physics Department, Autonomous University of Madrid (UAM), Spain*
[b] *Nicolás Cabrera University Institute of Materials Science (INC), UAM, Spain*
[c] *ALBA Synchrotron Light Facility, Cerdanyola del Vallès, Spain*
[d] *Norwegian University of Science and Technology (NTNU), Trondheim, Norway*
[e] *Fraunhofer Institute for Ceramic Technologies and Systems (IKTS), Dresden, Germany*
[f] *Instituto de Ciencia de Materiales de Madrid, Consejo Superior de Investigaciones Científicas (CSIC), Spain*
[g] *National Institute of Advanced Industrial Science and Technology (AIST), Research Institute for Energy Efficient Technologies, Japan*
[h] *Condensed Matter Physics Center (IFIMAC), UAM, Spain*
[*] *Corresponding authors: Celia Polop – celia.polop@uam.es*

## Abstract

Achieving reversible anode-free solid-state batteries hinges on controlling alkali-metal plating and stripping at buried interfaces, yet the underlying nanoscale mechanisms remain unresolved. Here we introduce virtual-electrode low-energy electron microscopy (VE-LEEM), an imaging platform that enables nanoscale visualization of anode formation and dissolution by combining electron-beam-induced plating with ultraviolet-driven stripping. By integrating VE-LEEM with synchrotron-based photoemission electron microscopy and atomic force microscopy, we track the chemical and morphological evolution of Li and Na anodes during cycling. We uncover a shared dynamic scaling regime governing anode growth, analogous to high-mobility thin-film deposition, but emerging through distinct morphological pathways dictated by metal surface energetics. In contrast, stripping proceeds through sequential grain-boundary unzipping and cluster-decay mechanisms, demonstrating that dissolution is intrinsically asymmetric with respect to plating and leaves behind a persistent interfacial residual layer. These results overturn the common assumption of mirrored plating–stripping behaviour and identify interfacial and grain-boundary energetics as fundamental constraints on reversibility. VE-LEEM thus provides a general route to resolve buried electrochemical interfaces at the nanoscale and establishes an energetic framework to guide the design of durable, high-energy anode-free solid-state batteries.

## Introduction

Achieving high-energy-density batteries for electric mobility requires breakthroughs beyond conventional lithium-ion technology. Solid-state batteries (SSBs) have emerged as leading candidates due to their potential for higher energy density and improved safety compared with liquid-electrolyte systems[1]. Among these, anode-free SSBs—also referred to as anode-less or zero-excess cells—offer additional advantages by eliminating the pre-assembled alkali-metal anode[2,3,4,5]. In this architecture, the cell is assembled in the discharged state, and the negative electrode (hereafter referred to as the anode) is electrochemically formed between the solid electrolyte and the current collector during charging. This approach eliminates electrochemically inactive alkali metal, thereby enabling the highest possible energy density for a given chemistry. Furthermore, anode-free SSBs offer significant advantages during cell assembly by avoiding direct handling of reactive metals, which typically requires inert atmospheres. This opens the prospect of air-based manufacturing and recycling, substantially lowering production and processing costs[6,7].

Despite these advantages, anode-free SSBs face critical challenges[8,9]. Importantly, these challenges are not specific to lithium chemistry, but arise from fundamental constraints governing alkali-metal plating at solid interfaces. The *in situ* formation of the alkali-metal anode requires injecting an incompressible metal with finite cohesive energy into the solid interface between electrolyte and current collector. The feasibility of this process depends on the chemical and mechanical properties of this interface, which must be thermodynamically stable, compliant, and ductile—yet strong enough to resist delamination, intrusion, and cracking. Maintaining conformal contact and mechanical integrity during plating and stripping is essential to avoid stress concentrations that trigger dendritic growth or dead-metal detachment. Because these constraints limit the reversibly cycled active material, performance is highly sensitive to active-material loss, directly impacting capacity retention and lifetime[10,11,12,13,14,15]. Accordingly, intense research efforts are devoted to stabilizing the electrolyte/anode/current-collector interface through lithiophilic coatings, pulsed-current protocols, engineered current collectors, hybrid electrolytes, and 3D host architectures[16,17,18]. Rational design of such interfaces, however, requires a fundamental understanding of the thermodynamics and kinetics governing the plating and stripping processes.

The nucleation and early anode growth stages are particularly critical, as they dictate the density, morphology, and reversibility of the plated metal. Numerous studies have investigated anode nucleation using optical and submicron electron microscopies[19,20]. However, the nuclei observed in these works—typically micrometer- to millimeter-scale features—are late-stage structures that obscure the fundamental nanoscale events governing plating and stripping. In contrast, true nuclei in the thermodynamic sense consist of only tens of atoms[21], which remain invisible to conventional probes. The gap between atomic-scale nucleation and macroscopic electrochemical performance has thus persisted as one of the most critical blind spots in the battery field [22,23,24,25]. Resolving this gap requires operando approaches capable of visualizing plating and stripping with true nanoscale resolution. As a result, the nanoscale events underpinning alkali-metal plating and stripping—such as wetting, smoothing, dendrite growth activation, and void formation—remain poorly understood.

Direct observation of anode formation at the buried solid-electrolyte/current-collector interface poses a major experimental challenge. This limitation can be overcome by using virtual electrodes (VEs): collimated, directional electron beams that act as electronically conductive current collectors, or charge injectors. A VE can charge the electrolyte surface, induce redox reactions, and drive ion migration while keeping the surface accessible to surface-sensitive probes. Mechanically, a VE behaves as an idealized, highly compliant current collector, maintaining perfect conformal contact with the anode/solid-electrolyte interface during cycling. As a result, the VE/anode interfacial energy approaches the surface energy of the alkali metal.

Recent 3D imaging studies have provided critical insight into the microstructure of anode-free SSBs under realistic mechanical constraints[19,20]. However, such approaches inherently probe systems in which metal growth is strongly influenced by physical current collectors and stack pressure. In contrast, the VE approach preserves electrochemical driving forces while decoupling them from mechanical constraints, enabling direct access to the nanoscale energetics governing plating and stripping. Previous implementations used low-energy electron guns coupled with non-spatially resolved X-ray photoelectron spectroscopy (XPS), or high-energy scanning electron microscopy (SEM), the latter posing a risk of solid-electrolyte damage[26,27,28].

Here, we introduce virtual-electrode low-energy electron microscopy (VE-LEEM), a technique that combines low-energy electron microscopy (LEEM) with ultraviolet (UV) illumination (Hg-lamp) to enable nanoscale imaging of plating and stripping in SSBs. In this configuration, the LEEM electron beam acts as a VE to induce metal plating while UV illumination drives stripping. By combining VE-LEEM with synchrotron-based photoemission electron microscopy (PEEM) and atomic force microscopy (AFM), we directly visualize, with nanoscale resolution, the dynamics of Na and Li anode formation and dissolution.

This work bridges the long-standing gap between nanoscale nucleation and macroscopic reversibility in anode-free SSBs. A central finding is that anode formation in both Na and Li follows a common dynamic scaling regime analogous to thin-film growth, but proceeds through distinct morphological pathways: Na exhibits strongly stochastic coalescence, whereas Li evolves more isotropically through a flooding-to-roughening transition governed by substrate topography. Stripping, by contrast, follows sequential regimes—high-energy grain-boundary unzipping and cluster decay—that diverge fundamentally from plating. These mechanisms challenge the long-standing assumption that stripping mirrors plating and identify nanoscale interfacial energetics as a major factor governing reversibility. Together, these findings establish an energetic framework for the intrinsic asymmetry between plating and stripping in anode-free SSBs and provide strategies for the rational design of durable, high-energy anode-free SSBs.

## Nanoscale visualization of plating and stripping in anode-free SSBs

We establish a reproducible, damage-free protocol to induce and image alkali-metal plating and stripping in anode-free SSBs, enabling direct nanoscale access to anode formation and dissolution at otherwise buried solid-state interfaces. The experimental workflow is illustrated schematically in Fig. 1. Polycrystalline $Na_5GdSi_4O_{12}$ (NaGdSiO) and single-crystal $Li_{6.5}La_3Zr_{1.5}Ta_{0.5}O_{12}$ (LLZTO) solid electrolytes were used as model systems for Na- and Li-based anode-free cells, respectively (Methods). In each case, a Na or Li metal foil was attached to the backside of the electrolyte and electrically connected, while the electrolyte surface remained accessible for imaging.

Plating and stripping were induced using VE-LEEM, in which the LEEM electron beam acts as an electronically conductive current collector. Both LEEM and PEEM operate in a cathode-lens configuration, applying a strong electric field between the sample surface and the objective lens (Fig. 1a). This geometry decouples the imaging voltage from the electron–surface interaction energy, allowing precise control of surface charging and redox reactions at electron energies below 20 eV. In contrast to conventional electron probes, this enables controlled electrochemical driving forces while avoiding beam-induced damage to the solid electrolyte.

Irradiation with low-energy electrons (6–15 eV) focused to a ~100 μm spot induces localized alkali-metal plating within well-defined surface regions (Fig. 1b). The negative surface charge supplied by the beam drives cation migration through the solid electrolyte to the surface, where ions are reduced and deposited as metallic Na or Li. Within this energy window, plating is homogeneous and reproducible, whereas higher electron energies (> 20 eV) lead to non-uniform growth and signs of electrolyte degradation (Extended Data Fig. S1). Because plating is spatially confined, multiple systematic experiments can be performed on a single sample by varying beam position, flux, energy, and deposited capacity.

Stripping is achieved by UV photoemission from a Hg lamp in combination with the PEEM extractor voltage. Electron removal generates localized positive charge at the surface, preferentially on the alkali metal due to its lower work function, driving ions back into the electrolyte and enabling controlled dissolution (Fig. 1b). Importantly, this process is reversible, allowing repeated plating–stripping cycles within the same region.

The resulting anodes were characterized *in situ* by synchrotron-based PEEM and *ex situ* by AFM (Fig. 1c). During PEEM measurements, photon flux and acquisition times were minimized to suppress unintended stripping or beam-induced damage. In PEEM, the choice between X-ray photoelectron spectroscopy (PEEM-XPS), X-ray absorption spectroscopy (PEEM-XAS), and UV photoelectron spectroscopy (PEEM-UV) modes to track chemical evolution during plating and stripping depends on the element of interest and the accessible photon-energy range of the beamline. Here, PEEM-XAS was applied to NaGdSiO, PEEM-XPS was preferred for LLZTO, and PEEM-UV was used in both systems. Next, the samples were transferred under an Ar atmosphere to a glovebox-integrated AFM for topographic imaging.

PEEM and AFM measurements were performed on different sample series prepared under identical cycling conditions; the images therefore provide representative rather than spatially identical views of a given anode state. Together, PEEM and AFM provide complementary chemical and morphological information on plated and stripped anodes at the nanoscale, enabling direct correlation between interfacial chemistry and growth dynamics.

## Contrasting lithium and sodium plating dynamics

Applying VE-LEEM reveals that, although Li and Na plate via similar nucleation-driven mechanisms, their growth diverges early into fundamentally different morphological pathways. Experiments were performed with an electron beam energy of 6 eV, using currents of 0.3 μA for Na and 1 μA for Li to ensure homogeneous, damage-free plating over practical timescales (Section S1 in the Supplementary Information). The comparison spans deposited capacities up to 100 $\mu Ah \cdot cm^{-2}$ for Na and 170 $\mu Ah \cdot cm^{-2}$ for Li.

For Na, PEEM–XAS images of the Na *K*-edge pre-peak ($h\nu$ = 1057 eV) reveal progressive cluster growth on NaGdSiO with increasing capacity (Fig. 2a–c). Dark regions correspond to Na clusters, clearly resolved in PEEM–XAS. AFM images (Fig. 2d–f) confirm discrete cluster nucleation and allow extraction of areal capacities from thickness analysis (Section S2 in the Supplementary Information). The near-linear evolution of plated capacity with time under constant beam conditions (Fig. S3) indicates that VE-LEEM realizes a well-defined local galvanostatic analogue, albeit implemented here as an electron-induced redox process under UHV conditions. Both PEEM and AFM indicate that Na plating initiates as isolated clusters that grow three-dimensionally (3D) and coalesce, developing irregular, fractal-like morphologies. Grain boundaries (GB) appear as bright, ramified contours in PEEM and as dark trenches in AFM. Spectral evolution at the Na *K*-edge shows a progressive dominance of the metallic Na feature at 1071.5 eV within plated regions, especially within clusters, accompanied by depletion of electrolyte-related peaks at higher energies (Fig. 2m)[29,30]. Spectral decomposition confirms that the plated regions correspond primarily to metallic Na (Fig. 2o, Extended Data Fig. S4).

Li plating on LLZTO proceeds through an analogous sequence of nucleation and 3D cluster growth, but subsequent coalescence yields markedly more compact and isotropic morphologies (Fig. 2g–l). In this system, lateral contrast between clusters and grain boundaries in PEEM–XPS is reduced due to charging effects in LLZTO, which broaden and shift core-level peaks. Reliable identification of plated regions relies on relative spectral evolution across growth stages rather than absolute binding energies[31]. Li 1s spectra collected over extended anode regions show increasing contributions from metallic Li at 53.5 eV with capacity, while electrolyte-related signals diminish (Fig. 2n). Spectral decomposition confirms increasing metallic Li with capacity and a diminishing electrolyte signal, while the intermediate component likely reflects a surface oxidation or solid-electrolyte interphase (SEI) (Fig. 2p and Extended Data Fig. S4).

Together, these observations demonstrate that both Na and Li plate through nucleation and 3D cluster growth, yet follow different coarsening pathways governed by distinct surface-energetic landscapes. The markedly higher surface energy of Li ($\gamma_s^{Li} \approx 0.5 - 0.6\ \mathrm{J{\cdot}m^{-2}}$) compared with Na ($\gamma_s^{Na} \approx 0.2 - 0.25\ \mathrm{J{\cdot}m^{-2}}$)[32] promotes compact, isotropic morphologies, whereas the lower surface energy of Na favors stochastic, fractal-like coalescence, highlighting surface energetics as a primary control parameter for anode morphology under anode-free conditions.

## Scaling laws governing anode-free metal plating

Beyond their distinct morphologies, Li and Na anodes exhibit remarkably similar scaling behavior during plating, revealing a common growth regime governed by dynamic scaling laws. Quantitative analysis of anode morphology shows that alkali-metal plating in anode-free SSBs follows scaling relationships analogous to those established for thin-film growth under high atomic mobility[21,33,34].

Morphological parameters were extracted from AFM measurements as a function of plated capacity $Q$ for both Na and Li (Fig. 3a,b). We track the anode roughness ($\rho_{anode}$, blue) and a characteristic cluster size ($\lambda$, red), where $\rho_{anode}$ is obtained from the measured surface roughness $\rho_{surf}$ after subtracting the substrate contribution, and $\lambda$ is determined from autocorrelation analysis and direct cluster counting (Section S5 in the Supplementary Information). For both systems, $\rho_{anode}$ and $\lambda$ increase concurrently with $Q$, consistent with progressive cluster coarsening. Experiments were conducted at homologous temperatures $T/T_m$ of ~0.8 for Na and ~0.65 for Li, which are conditions of enhanced atomic mobility, where multiple diffusion pathways are activated.

Both roughness and cluster size follow clear power-law scaling with capacity, $\rho_{anode} \sim Q^{\beta}$ and $\lambda \sim Q^{1/z}$. For Na, we obtain $\beta$ = 0.9 ± 0.1 and $1/z$ = 1.1 ± 0.1, while for Li $\beta$ = 1.2 ± 0.1 and $1/z$ = 1.0 ± 0.1. Within experimental uncertainty, the scaling exponents are comparable for the two systems, clustering around $\beta \approx 1$ and $1/z \approx 1$.

Within the framework of dynamic scaling theory developed for thin-film growth, a growth exponent $\beta > 0.5$ (random deposition corresponds to $\beta = 0.5$) indicates uphill mass transport driven by thermodynamic forces[33]. At high homologous temperatures, this regime reflects surface- and interface-energy minimization through dewetting, faceting, and curvature-driven diffusion rather than kinetically limited deposition. Notably, alkali metals at room temperature operate at homologous temperatures > 0.6, in contrast to most thin-film growth systems, such that surface diffusion, GB diffusion and creep are simultaneously active. The observed exponents therefore indicate that alkali-metal plating proceeds under near-equilibrium growth conditions.

Consistent with this interpretation, the anode roughness scales proportionally with cluster height ($\rho_{anode} \propto h_{cluster}$; Fig. S6a,b), yielding a cluster aspect ratio $AR_{cluster} = h_{cluster}/\lambda \sim Q^{\beta-1/z}$ that approaches a constant value during plating (Extended Data Fig. S6c,d). This behavior indicates that clusters evolve toward self-similar, pseudo-equilibrium shapes that persist as deposition proceeds.

Together, these results establish that anode-free Na and Li plating obey shared dynamic scaling laws, consistent with thin-film growth in the high-mobility regime. Despite their distinct surface energies and coarsening pathways, both systems converge toward self-similar growth regimes governed by the same energetic principles. On this basis, the following sections focus on Li to examine the earliest stages of plating, and on Na to elucidate stripping dynamics.

## Early-stage instabilities in lithium plating

We capture the emergence of the first Li nuclei and show that the early stages of Li plating are governed by the nanoscale morphology of the solid electrolyte surface, which controls where and how the first nuclei emerge (Fig. 4). All experiments were conducted using a 6 eV, 1 μA electron beam. The LLZTO surface exhibits substantial roughness ($\rho_{surf}$ = 19 ± 1 nm), arising from two distinct topographic components: large mounds (~150 nm; magenta in Fig. 4a) superimposed on a finer-grained background (~45 nm; cyan). Height-distribution analysis (Fig. S5) shows that this background contributes $\rho_{electrolyte_bg}$ = 9.2 nm to the total surface roughness.

At the earliest stages of deposition (Fig. 4b), Li accumulates preferentially within the fine-grained background topography, effectively flooding the electrolyte surface. The average background grain size increases to 64 ± 4 nm (green), while the overall surface roughness decreases to 15 ± 1 nm at a plated capacity of 7 μAh cm$^{-2}$. Taller features (blue) persist as remnants of the original LLZTO mounds. Upon further deposition (≥ 11 μAh cm$^{-2}$; Fig. 4c,d), the electrolyte surface becomes fully covered and rounded Li clusters (black) emerge and coarsen. This transition is reflected in the height profiles (Fig. 4e–h).

The initial flooding regime explains the deviation from power-law scaling observed for the first two data points in the roughness evolution (enclosed by a dashed black line in Fig. 3b). Once the electrolyte is buried, subsequent deposition leads to surface roughening. Slope distributions (Fig. 4i) and the evolution of the mean surface slope $m$ (Fig. 4j) clearly distinguish these two regimes: a flooding regime characterized by decreasing roughness and slope, followed by a roughening regime marked by their concurrent increase.

Cluster-scale analysis reveals that Li clusters form through coalescence of grains meeting along inner GBs with high dihedral angles ($\phi_{in}$ = 150 ± 10°), characteristic of low-energy boundaries between slightly misoriented grains (dashed grey lines in Fig. 4c,d). Growth proceeds primarily through internal grain coarsening while preserving overall cluster shape. Consistent with the scaling analysis, the characteristic cluster size $\lambda \sim Q^{1/z}$ captures grain coalescence mediated by low-energy GBs. In contrast, steep outer GBs (dashed white lines) form at cluster interfaces with nearly constant dihedral angles ($\phi_{out}$ = 91 ± 7°; Fig. S6f), indicative of high-energy boundaries that suppress lateral coalescence and promote vertical growth.

This transition from flooding to roughening can be rationalized by the balance of surface and interfacial energies during plating. Initially, Li wets the rough LLZTO surface because the electrolyte surface energy $\gamma_S^{LLZTO}$ (≈ 0.9–1.4 J·m$^{-2}$, depending on surface termination and chemical environment) exceeds the sum of the Li surface energy $\gamma_S^{Li}$ (≈ 0.5 J·m$^{-2}$) and the electrolyte/anode interfacial energy $\gamma_i^{LLZTO/Li}$ (≈ 0.2–0.6 J·m$^{-2}$), consistent with density functional theory estimates and literature values[32,35]. Once the electrolyte surface is fully flooded, further growth is governed by the competition between surface and GB energies. When the combined grain-boundary and elastic energy ($\gamma_{GB} + \gamma_e$) falls below an effective coincidence-site lattice (CSL)-like energy scale $\gamma_{\mathrm{GB}}^{\mathrm{CSL}}$, low-energy inner GBs can form via grain reorientation and facilitate coalescence[36]. From the slope evolution, we estimate $\gamma_{GB}^{CSL} + \gamma_e \approx 0.24$ J·m$^{-2}$ (≈ 0.52· $\gamma_s^{Li}$) for slightly misoriented grains, whereas highly misoriented outer GBs exhibit $\gamma_{GB} + \gamma_e \approx 0.57$ J·m$^{-2}$ (≈ 1.23·$\gamma_s^{Li}$) (Section S7 in the Supplementary Information).

Using VE-LEEM in combination with PEEM and AFM, we thus directly resolve the transition from substrate flooding to cluster coarsening and roughening (schematic in Fig. 4k). These results demonstrate that nanoscale electrolyte roughness critically controls Li nucleation and early growth. Once nucleation is established, subsequent plating is governed by the balance of surface and interfacial energies, driving the evolution from conformal flooding to roughened, near-equilibrium morphologies. More broadly, these findings show that early-stage instabilities in anode-free Li plating are determined not only by electrochemical conditions, but also by the nanoscale energetic landscape of the electrolyte interface.

## Non-uniform stripping dynamics in sodium anodes

Stripping of Na anodes does not retrace the plating pathway but instead proceeds through sequential dissolution mechanisms that diverge fundamentally from growth (Fig. 5). To probe these dynamics, stripping was initiated from fully plated Na anodes (79 µAh·cm$^{-2}$) with an initial surface roughness of $\rho_{surf}$ = 190 nm, characteristic cluster size $\lambda$ = 2.5 µm, and average cluster height $h_{cluster}$ = 800 nm (Fig. 5a).

AFM images acquired after 14, 42, and 70 min of UV-induced stripping reveal a highly non-uniform morphological evolution (Fig. 5b–d). The analysis of the evolution of the stripped capacity with UV exposure time is shown in the SI (Section S3). Height profiles (Fig. 5e–h) show that during the initial stages of stripping, deep trenches form and propagate toward the electrolyte surface as high-energy outer GBs unzip (magenta arrows in Fig. 5f), while the average cluster height remains nearly constant. This process leads to a transient increase in surface roughness to $\rho_{surf}$ ≈ 240 nm. At later stages, both $h_{cluster}$ and $\lambda$ decrease concurrently, signaling a transition to a distinct dissolution regime.

The evolution of $h_{cluster}$, $\lambda$, and the cluster-edge slope $m_{edge}$ as a function of stripped capacity (Fig. 5i) reveals two stripping regimes. In the first regime, outer-GB unzipping, the lateral cluster size decreases slightly faster ($\lambda \sim t^{-0.4}$) than the cluster height ($h_{cluster} \sim t^{-0.3}$), while $m_{edge}$ increases, consistent with preferential removal of high-energy boundaries. In the second regime, cluster decay, the cluster height collapses rapidly ($h_{cluster} \sim t^{-3.5}$) compared with the lateral size ($\lambda \sim t^{-2.5}$), accompanied by a decrease in $m_{edge}$, indicating a transition toward more isotropic dissolution.

Even after extended stripping (>70 min), a thin residual layer persists at the electrolyte interface (<1 μAh cm$^{-2}$; Fig. 5d). This irreversible residual layer is summarized schematically in Fig. 5j, together with the two stripping regimes, and contrasted with the full plating–stripping cycle sketched in Fig. 5k, which highlights the distinct trajectories of roughness and cluster size during growth and dissolution.

These observations demonstrate that Na stripping proceeds through sequential, non-mirror mechanisms fundamentally distinct from plating. Initial stripping is driven by the removal of high-energy outer GBs, reflecting a thermodynamic preference for eliminating interfaces with large energy penalties. Once these boundaries are exhausted, dissolution proceeds through cluster shrinking governed primarily by surface-energy minimization. The persistence of an irreversible residual layer at the anode–electrolyte interface represents an intrinsic limit to reversibility in anode-free Na SSBs, with direct implications for capacity retention and cycling stability.

## Cycling reversibility and mechanistic implications

To evaluate the reversibility of alkali-metal cycling under VE-LEEM conditions, we performed sequential plating and stripping experiments over four consecutive cycles using identical parameters, followed by *in situ* PEEM-UV and *ex situ* AFM characterization after each cycle. The purpose of these experiments is to elucidate the nanoscale mechanisms governing reversibility and morphological evolution, rather than to assess long-term cyclability or device lifetime. Figures 6a–f show the surface morphology of the Na anode after 0.5, 1.5, and 3.5 cycles. Here, 0.5 cycles correspond to the initial plating discussed above, while 1.5 cycles denote a full plating–stripping–plating sequence, and so forth.

Across the first four cycles, the overall anode morphology appears largely reversible, with the surface recovering its main structural features after each plating step. Notably, no progressive roughening, crack formation, or morphological signatures of electrolyte degradation are observed, indicating that repeated VE-LEEM cycling does not induce detectable damage to the solid electrolyte or the anode–electrolyte interface under the conditions employed.

Quantitative analysis of areal capacity ($Q$), anode roughness ($\rho_{anode}$), and characteristic cluster size ($\lambda$) throughout cycling is shown in Fig. 6j,k. The cluster size $\lambda$ remains constant after each plating step, demonstrating reproducible nucleation across cycles. The most pronounced deviation from perfect reversibility occurs after the first stripping event: during the second plating step, the overall anode roughness $\rho_{anode}$ is slightly reduced relative to the initial cycle. This change is attributed to the thin residual layer left behind after the first stripping cycle, which modifies local surface and interfacial energetics and, in turn, influences subsequent grain growth. The data points at cycle numbers 2 and 3 in Fig. 6j,k are extrapolated and set equal to the values after one full cycle, as the capacity and anode roughness returns to approximately zero after a complete cycle within experimental uncertainty; further AFM measurements on almost empty surfaces were therefore not performed.

Consistent with this interpretation, AFM height profiles reveal that trenches associated with GBs become shallower during later plating steps (magenta arrows in Fig. 6h,i), reflecting partial smoothing of the energetic landscape rather than irreversible morphological degradation. Importantly, this evolution does not result in measurable capacity loss over the cycles investigated, indicating that the residual layer alters morphology without compromising electrochemical reversibility under VE-LEEM conditions.

Together, these results demonstrate that Na anodes retain a high degree of morphological reversibility over multiple plating–stripping cycles, despite the persistence of an irreversible residual layer. This finding highlights two key mechanistic implications. First, it confirms that VE-LEEM enables repeated, damage-free cycling, in contrast to VE probes based on conventional electron beams, which often induce beam damage or electrolyte degradation. Second, it shows that early-stage stripping-induced modifications of interfacial energetics can subtly redirect subsequent growth pathways without immediately limiting reversible capacity. These observations bridge the nanoscale stripping mechanisms identified above with their cumulative impact on cycling behavior, providing a mechanistic link between interfacial energetics and anode reversibility.

## Discussion

By directly visualizing alkali-metal plating and stripping with nanoscale resolution, this work provides an integrated mechanistic framework linking interfacial energetics, microstructural evolution, and reversibility in anode-free SSBs. Using a VE approach combined with complementary spectroscopic and topographic probes, we show that anode formation converges toward a common growth regime once a continuous metal layer is established, whereas dissolution proceeds through fundamentally asymmetric pathways governed by grain-boundary and surface energetics. Together, these mechanisms help rationalize previously disparate observations across the literature and highlight energetic constraints that can intrinsically limit reversibility in anode-free architectures.

**Common dynamic scaling in anode-free Li and Na plating**

A central finding is that both Li and Na plating rapidly converges toward shared dynamic scaling laws characteristic of thin-film growth under high atomic mobility. The power-law scaling of roughness and characteristic cluster size with deposited capacity (Fig. 3) matches classical growth models, indicating that alkali-metal plating is predominantly governed by thermodynamic driving forces rather than kinetically limited deposition once growth is established. Previous *ex situ* studies reported microstructural heterogeneity but lacked access to the temporal evolution of growth parameters and therefore could not identify the underlying scaling regime[19,22,37]. Importantly, observing the same dynamic-scaling behaviour in two technologically central anode-free systems (Li and Na) suggests that scaling analysis can serve as a broadly transferable framework to quantify and compare plating pathways across materials and interfaces, motivating systematic tests of its universality under different solid electrolytes, interphases and mechanical constraints.

This shared scaling behavior does not emerge from the earliest stages of deposition. For Li, we resolve a transition from substrate flooding to surface roughening governed by the nanoscale topography of the solid electrolyte (Fig. 4). Early nucleation heterogeneities associated with fine-grained electrolyte features dominate the flooding regime but are progressively screened once the surface is fully buried. The common growth regime is reached at very small deposited amounts (~10 μAh $cm^{-2}$, corresponding to ~30 nm of equivalent metal thickness), well below the thickness scales typically considered in battery studies. These observations help rationalize apparently conflicting reports on preferential Li nucleation at GBs or defects, suggesting that such preferences may be most prominent during the earliest stages of growth and become progressively screened once a continuous metal layer forms, with the subsequent evolution converging toward self-similar behaviour under near-equilibrium growth conditions[28].

**Plating–stripping asymmetry as an energetic inevitability**

While plating converges toward a common growth regime, stripping follows a fundamentally different trajectory. For Na, dissolution proceeds through two sequential, non-mirror mechanisms: initial unzipping of high-energy outer GBs, followed by isotropic cluster decay driven by surface-energy minimization (Fig. 5). This behavior challenges the common assumption that stripping is the time-reversed counterpart of plating and provides a mechanistic origin for asymmetric electrochemical signatures reported in anode-free and metal-anode SSBs[17,38], which previously lacked direct links to specific grain-boundary and interfacial energetics.

A direct consequence of this asymmetry is the persistence of a thin interfacial layer after extended stripping. Although stripping is induced here by UV photoemission under UHV conditions, the persistence of residual interfacial material suggests an energetic stabilization under the conditions studied here, rather than a modality-specific artefact. Similar interfacial residues have been inferred from Coulombic inefficiency and post-mortem imaging in electrochemically stripped anode-free cells, supporting the broader relevance of this phenomenon. While the present data do not yet resolve the precise chemical state of this residual layer, its persistence points to an energetic stabilization at the metal–electrolyte interface, imposing an intrinsic limit on reversible capacity in anode-free SSBs. Future chemically resolved operando and spectromicroscopy experiments will be required to determine the chemical nature of this residual interfacial layer.

**Cohesive limitations of the alkali-metal anode**

Beyond interfacial effects, cohesive limitations within the alkali-metal anode constitute an additional intrinsic constraint on reversibility. The plated anodes exhibit polycrystalline morphologies fragmented by GBs and trenches that act both as preferred growth sites and as dissolution pathways, concentrating stress and facilitating void and pore formation during cycling. Similar features have been observed *ex situ* under stack pressure[9,39], but without access to their nanoscale dynamics. Given that Li and Na operate at homologous temperatures exceeding 0.6 $T_m$, even modest stress gradients can activate creep and diffusion-assisted relaxation. The preferential localization of dissolution along GB trenches is consistent with stress- and cohesion-related instabilities within the metal anode.

Cycling under VE-LEEM further shows that, although gross morphology remains largely reversible (Fig. 6), early stripping-induced modifications to interfacial energetics subtly redirect subsequent growth pathways. The persistence of the interfacial residual layer alters GB topology without immediate capacity loss, highlighting that morphological reversibility does not imply complete energetic reversibility. Together with the stripping asymmetry, these results indicate that both interfacial energetics and cohesive limitations act as intrinsic constraints on reversibility in anode-free SSBs.

**Implications for anode-free battery design**

The mechanisms identified here imply that stabilizing the anode-electrolyte interface alone is insufficient if the alkali-metal anode remains structurally fragmented. Reducing GB density— by promoting more uniform growth or enhancing surface diffusivity—thus emerges as an key design objective. Ultrathin alloyable interlayers or surfactant-like interphases that homogenize interfacial energies over relevant diffusion length scales may promote uniform anode formation without increasing operating temperature, which is already close to the melting regime for both Li and Na. At the same time, achieving a chemically and mechanically conformal interface with the current collector remains critical but cannot alone ensure dense and reversible anode formation, as mechanical mismatch and localized stress concentration promote delamination and void formation[1,40]. Effective design strategies must therefore address both interfacial energetics and cohesive integrity, guided by nanoscale mechanisms rather than macroscopic heuristics.

**Scope and outlook**

VE-LEEM probes model solid electrolytes under UHV conditions with localized, charge-controlled current injection, intentionally excluding redox-driven interphase formation and space-charge effects present in practical devices. The mechanisms identified here therefore reflect intrinsic interfacial and cohesive energetics rather than full-cell behaviour under realistic stack pressure and current densities. Although VE-LEEM decouples electrochemical driving forces from stack pressure and realistic cell architectures, the energetic hierarchies resolved here are expected to govern the earliest stages of anode formation in practical cells, where irreversible losses are seeded. In anode-free configurations, the first tens of nanometres of metal deposited at the solid-electrolyte/current-collector interface critically determine subsequent morphology, contact stability and reversibility. The nanoscale mechanisms identified here are therefore expected to be most relevant during the initial nucleation and early growth stages, before mechanical confinement and stack pressure dominate the evolution of the anode microstructure.

The energetic trends identified here originate from fundamental surface and grain-boundary thermodynamics and are therefore expected to be broadly transferable across different solid electrolytes and anode-free architectures. VE-LEEM thus provides a versatile platform for resolving buried interfacial mechanisms and informing the rational design of durable, high-energy anode-free solid-state batteries.

## Conclusion

In conclusion, this work provides direct nanoscale insight into alkali-metal plating and stripping in anode-free solid-state batteries. By combining virtual-electrode low-energy electron microscopy (VE-LEEM) with complementary spectroscopic and topographic probes, we show that anode formation follows a shared dynamic-scaling regime in Li and Na once a continuous metal layer is established, whereas dissolution proceeds through asymmetric, non-mirror pathways governed by interfacial and grain-boundary energetics. Early-stage instabilities in lithium plating originate from nanoscale electrolyte topography, whereas sodium stripping proceeds through sequential grain-boundary unzipping, cluster decay and the formation of a persistent residual layer. The observation of shared scaling behaviour in both Li and Na points to a general, quantitative framework to compare anode-free plating across chemistries, enabling systematic scaling-based benchmarks under more realistic interfaces and mechanical constraints.

These findings establish an energetic framework linking nanoscale morphology evolution to macroscopic reversibility, helping to rationalize why capacity loss in anode-free architectures can have an intrinsic energetic component rather than being purely protocol-dependent. More broadly, our results indicate that controlling surface, interfacial and cohesive energetics at the nanoscale is central to enabling durable, high-energy anode-free solid-state batteries. The VE-LEEM approach introduced here provides a general route to systematically probe buried solid-state interfaces that have remained inaccessible to conventional experimental techniques.

## Methods

### Solid electrolyte preparation

Polycrystalline $Na_5GdSi_4O_{12}$ (NaGdSiO) was prepared by melt quenching followed by controlled crystallization. Stoichiometric amounts of $Na_2CO_3$ (VWR, AnalaR NORMAPUR), $SiO_2$ (Millisil W8, Quarzwerke), and $Gd_2O_3$ (Alfa Aesar) were homogenized in a Turbula mixer for 30 min and melted in a platinum crucible at 1350 °C for 1 h. The melt was quenched on a brass block to form a glass frit, which was subsequently pre-milled using a disc mill (Retsch RS1) and ball-milled for 12 h (Fritsch Pulverisette 5). Crystallization was induced by calcination in air at 950 °C for 1 h, yielding the conductive $Na_5GdSi_4O_{12}$ phase. The powder was further refined by attrition milling (Netzsch, 1000 rpm, 1 h), following established procedures[41]. NaGdSiO pellets (8 mm diameter, ~1.3 mm thickness) were fabricated by uniaxial pressing (0.25 g) and sintered in air at 1120 °C for 2 h. After sintering, the pellets were transferred to an argon-filled glovebox and annealed at 700 °C for 4 h to remove surface contamination from air exposure. The surface intended for VE-LEEM was polished using SiC abrasive papers down to P4000 grit. The opposite surface was roughened with P320 grit, contacted with sodium metal foil, and hot-pressed at 97 °C under 10 MPa for 5 min to ensure electrical contact.

Single-crystal $Li_{6.5}La_3Zr_{1.5}Ta_{0.5}O_{12}$ (LLZTO) was grown by a melting-based crystal growth method following ref.[42]. Samples were either cut into cuboids (6 × 6 × 1 $mm^3$) or used in their as-grown circular form (Ø 11 mm, ~0.7 mm thickness). The electron beam-facing surfaces were polished with an automated polishing system (Tegramin-20) using successive SiC papers (P1200–P4000), followed by diamond suspensions (3, 1, and 0.25 µm). To remove surface carbonate layers, samples were immersed in 0.1 M HCl for 1 min, rinsed with isopropanol, and dried. One side of each crystal was then sputter-coated with Au (Edwards S150B, 20 kV, 3 min) to promote wetting by Li metal. After coating, the opposite surface underwent a second acid treatment and immediately transferred into an argon-filled glovebox ($O_2$, $H_2O$ < 1 ppm). Inside the glovebox, Li metal was melted and applied onto the Au-coated side to form the counter electrode.

**Low energy electron microscopy and photoemission electron microscopy**

LEEM and PEEM experiments were performed using an Elmitec LEEM/PEEM instrument operated in cathode-lens configuration. In this geometry, electrons are accelerated to high voltage (10–20 kV) and decelerated close to the sample surface, allowing precise control of the landing energy in the range 0–200 eV while maintaining high spatial resolution. All experiments were conducted at an imaging voltage of 20 kV with a working distance of 4–5 mm.

The incident electron kinetic energy at the sample surface is controlled by the start voltage applied to the sample stage. Sample charging acts as an effective offset to this voltage and was monitored in situ via the mirror electron microscopy (MEM)–to–LEEM transition and shifts in UV photoemission thresholds. Owing to the ionic conductivity of the solid electrolytes and electrical contact at the backside, surface charging did not exceed a few volts under the conditions employed.

In VE-LEEM experiments, the low-energy electron beam serves as a virtual electrode, supplying negative charge locally to the electrolyte surface and driving cation migration and reduction at the interface. Importantly, the relevant experimental control parameter is the total injected charge and is quantified independently by AFM-derived areal capacity. Under UHV conditions, this process constitutes an electron-induced redox reaction without competing chemical reduction pathways, providing a well-defined analogue to electrochemical plating. This enables comparison of growth trajectories at fixed deposited capacity without conversion to device-scale current densities. The linear time-dependence of capacity is consistent with a local galvanostatic cycling (Section S3 in the Supplementary Information).

PEEM measurements were performed using ultraviolet illumination or synchrotron radiation. Energy-filtered PEEM enabled spatially resolved X-ray photoelectron spectroscopy (XPS) and X-ray absorption spectroscopy (XAS), combining chemical sensitivity with lateral resolution on the tens-of-nanometers scale. While both techniques provide lateral resolution down to the tens of nanometers, the higher intensity in XAS mode translates into faster acquisition and slightly improved effective spatial resolution under practical conditions. The probing depth is limited by the electron escape depth and is ~5-10 nm in XAS and smaller in XPS (~1 nm). PEEM-XAS was used preferentially for NaGdSiO owing to its higher signal intensity, while PEEM-XPS was employed for LLZTO to track Li chemical states. The field of view (FOV) of PEEM images is nominal and can vary slightly with focusing conditions.

### Atomic force microscopy

Atomic force microscopy (AFM) was performed using a Park Systems FX40 microscope mounted on an Accurion i4 active vibration isolation platform, ensuring vertical noise levels of ~ 2 ± 1 Å. The system was installed inside an MBraun LABmaster Pro ECO glovebox operating under Ar atmosphere ($O_2$, $H_2O$ < 0.1 ppm).

AFM images of the Li and Na anodes ranged from 1 × 1 $\mu m^2$ to 60 × 60 $\mu m^2$, and were taken in amplitude-modulation AFM using PPP-NCHR (Nanosensors) and AC160TS (Olympus) cantilevers with nominal resonance frequencies of 330 and 300 kHz respectively and tip radii below 10 nm. Higher-resolution images (1 × 1 $\mu m^2$) were acquired using SSS-NCHR probes with tip radii below 2 nm. AFM length scales were calibrated using reference standards. Data processing and quantitative analysis were performed using WSxM 5.0[43]. PEEM and AFM data shown in this work were acquired on different sample series prepared under identical cycling conditions and are therefore not spatially correlated one-to-one.

## Acknowledgements

This work was supported by the European Union's Horizon Europe Research and Innovation Programme under Grant Agreement No. 101103834 (OPERA project), part of the Battery2030+ initiative. The views and opinions expressed are those of the authors only and do not necessarily reflect those of the European Union or CINEA. Neither the European Union nor the granting authority can be held responsible for them. We gratefully acknowledge access to beamline CIRCE at the ALBA Synchrotron. We also acknowledge support from the Spanish Ministry of Science, Innovation and Universities under Grants PCI2022-132998 and PID2021-1246670B, and through the "María de Maeztu" Programme for Units of Excellence in R&D (CEX2023-001316-M).

## Author contributions

E.G.M., M.A.N., M.F. and C.P. conceived the project and the underlying ideas. R.A., A.L. and K.N. prepared the NaGdSiO samples. H.R.B., K.K. and D.R. prepared the LLZTO samples. J.D.S., P.H.M., N.K.M., A.G., M.A.N. and M.F. performed and analyzed the synchrotron-based experiments. J.D.S. and P.H.M. performed and analyzed the AFM experiments. C.P. led the analysis and interpretation, aided by J.D.S. E.V. and C.P. developed the mechanistic framework. All authors discussed the results. C.P. wrote the manuscript with input from all authors.

## Competing interests

The authors declare no competing interests.

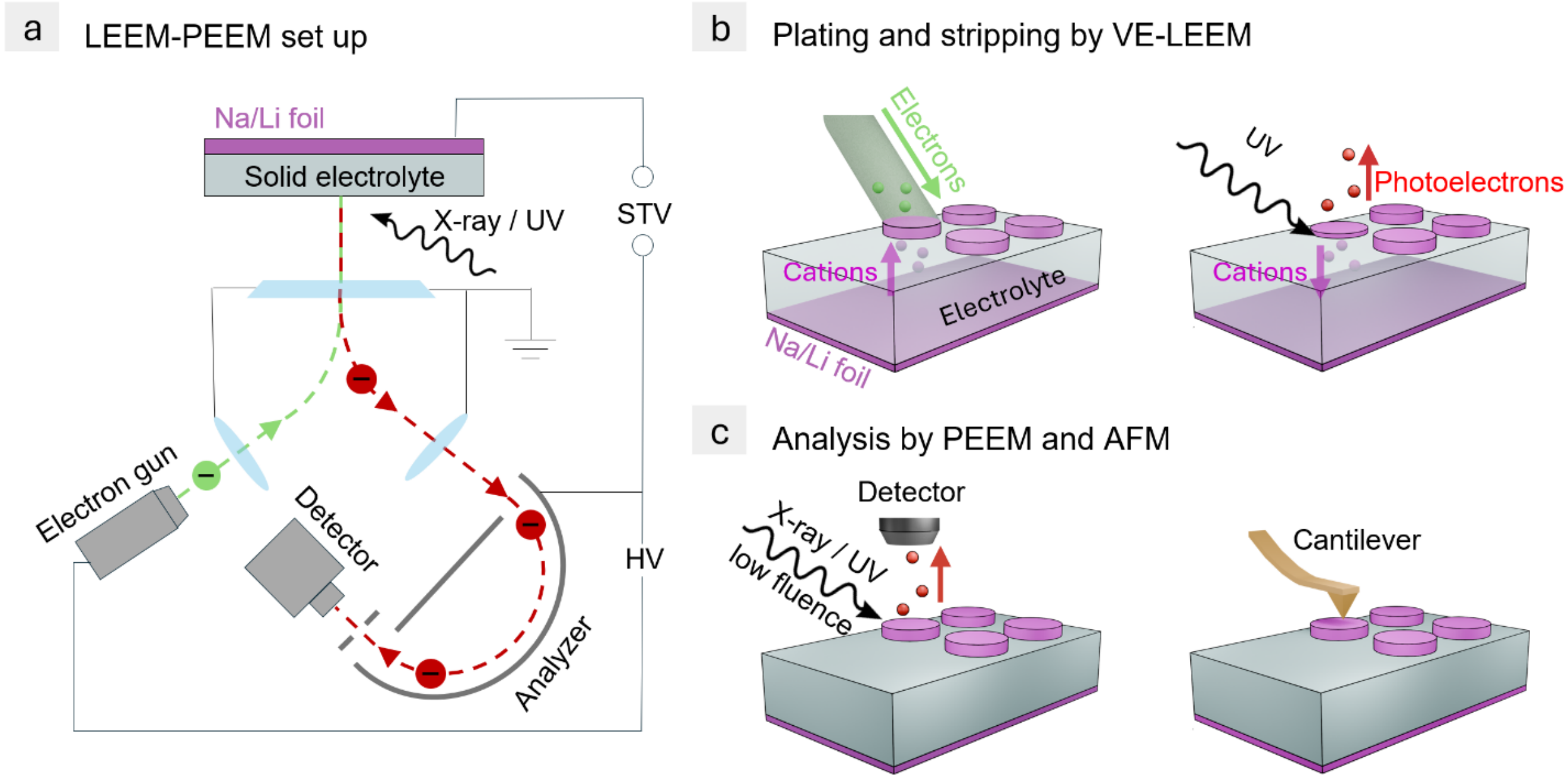

**Figure 1 | The VE-LEEM platform.**

a, Schematic of the VE-LEEM setup combining low-energy electron microscopy (LEEM) and photoemission electron microscopy (PEEM) for imaging of plating and stripping at anode-free solid-state batteries.

b, Localized plating and stripping induced by the LEEM electron beam and ultraviolet illumination (purple circles), demonstrating that electron injection and photoemission enable reversible, spatially confined redox reactions.

c, Workflow for integrated VE-LEEM, synchrotron-based PEEM, and ex situ AFM characterization, enabling correlative chemical and morphological analysis of plating and stripping dynamics with nanoscale resolution.

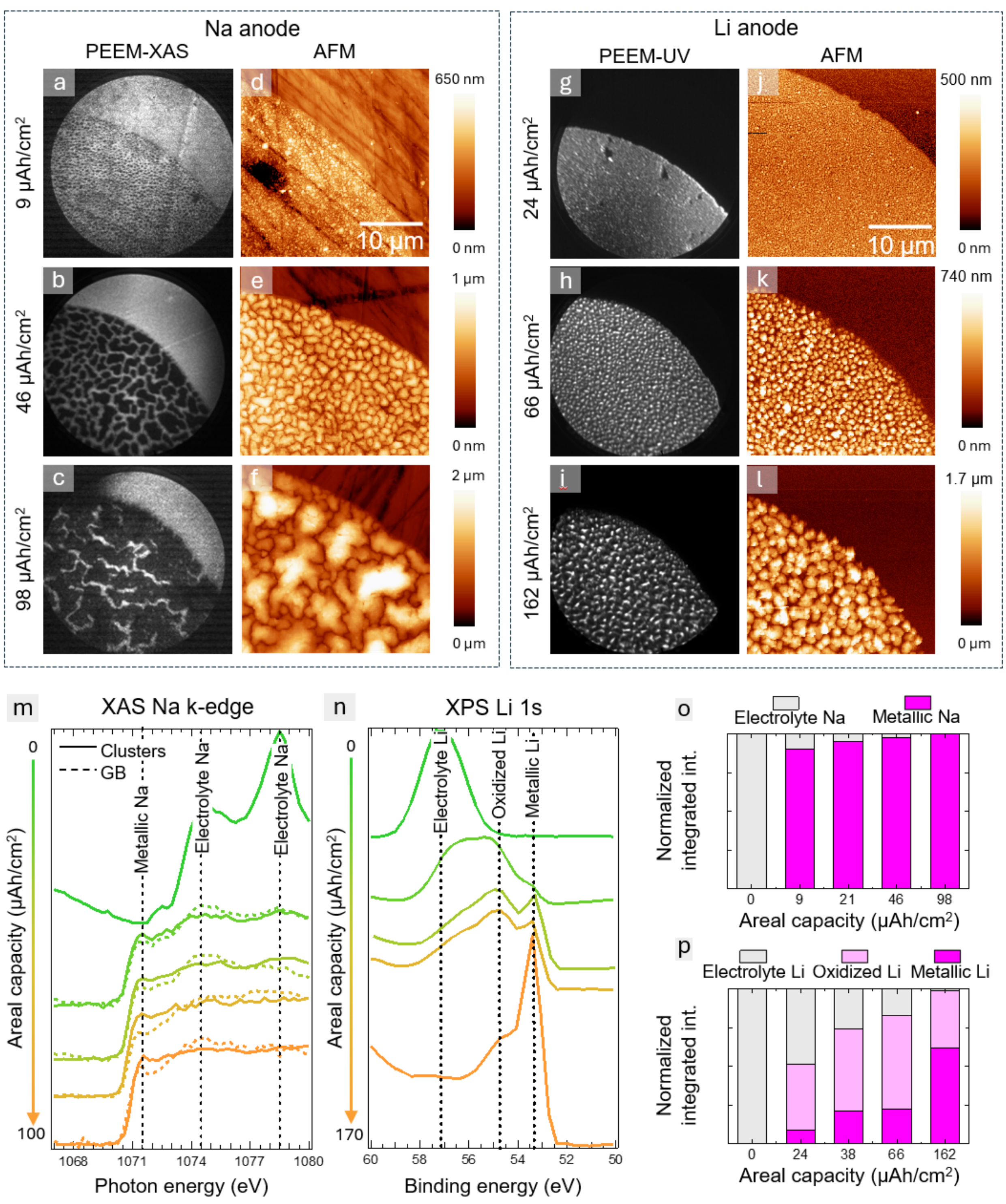

**Figure 2 | Contrasting Na and Li plating.**

a–c, PEEM-XAS images of Na plating on NaGdSiO showing cluster nucleation, coalescence, and fractal-like growth with increasing capacity. Nominal FOV = 30 µm.

d–f, Corresponding AFM topographies revealing discrete three-dimensional clusters and grain-boundary trenches.

g–l, Li plating on LLZTO exhibits more compact and isotropic morphologies at comparable deposited capacity. Nominal PEEM FOV = 30 μm.

m–p, Evolution of Na K-edge and Li 1s spectra with increasing capacity, showing the progressive emergence of metallic components and depletion of electrolyte-related features.

Both Na and Li plate via three-dimensional cluster growth, yet follow distinct coarsening pathways governed by differences in surface energetics.

PEEM and AFM images correspond to representative regions obtained under identical plating conditions and are not taken from the exact same spot. Small differences in apparent cluster size between PEEM and AFM images primarily reflect the approximate spatial calibration of PEEM, whose real field of view can vary slightly with focus conditions, whereas AFM length scales are calibrated against reference standards.

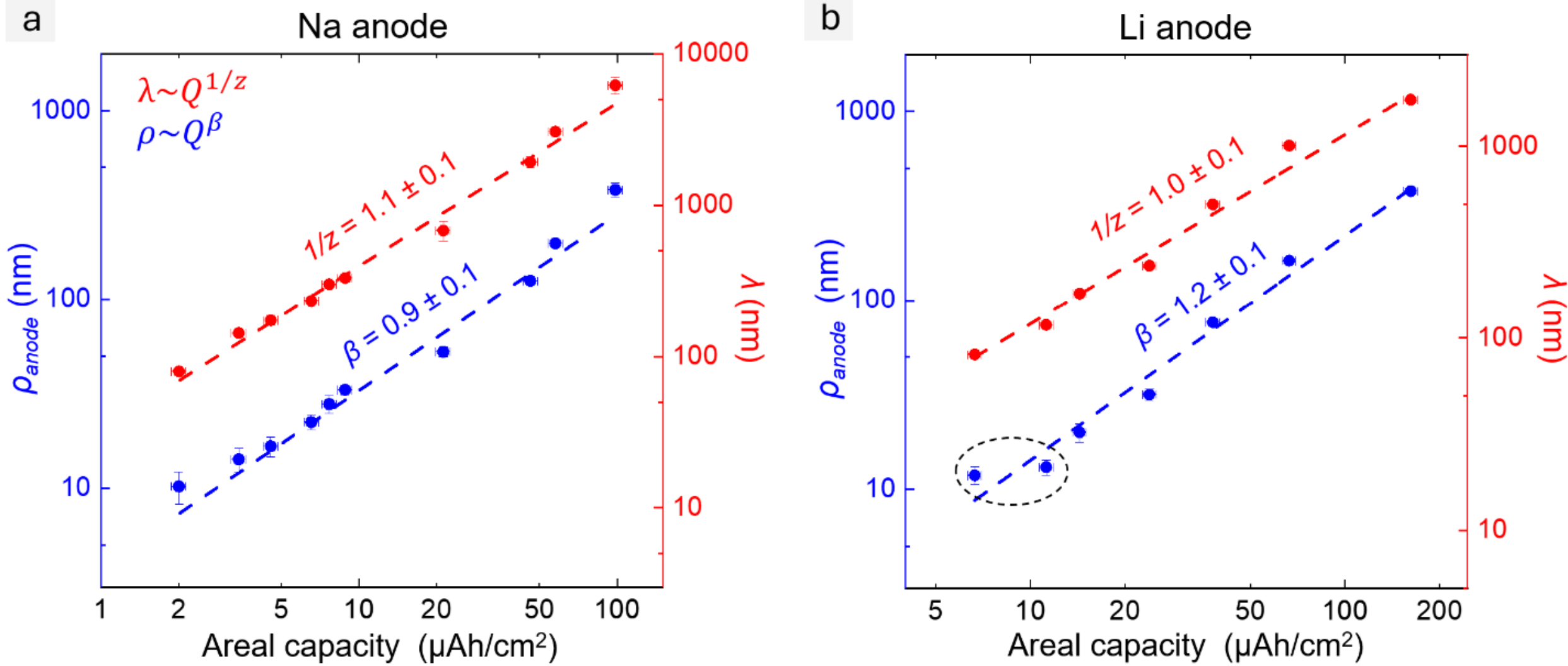

**Figure 3 | Scaling laws governing anode plating.**

a,b, Evolution of anode roughness ($\rho_{anode}$, blue) and characteristic cluster size ($\lambda$, red) as a function of plated capacity $Q$ for Na and Li. Both parameters exhibit power-law scaling ($\rho_{anode} \sim Q^{\beta}$, $\lambda \sim Q^{1/z}$) with exponents $\beta \approx 1$ and $1/z \approx 1$. These exponents indicate near-equilibrium growth governed by thermodynamic driving forces rather than kinetically limited random deposition, revealing shared dynamic scaling behaviour across different alkali metals.

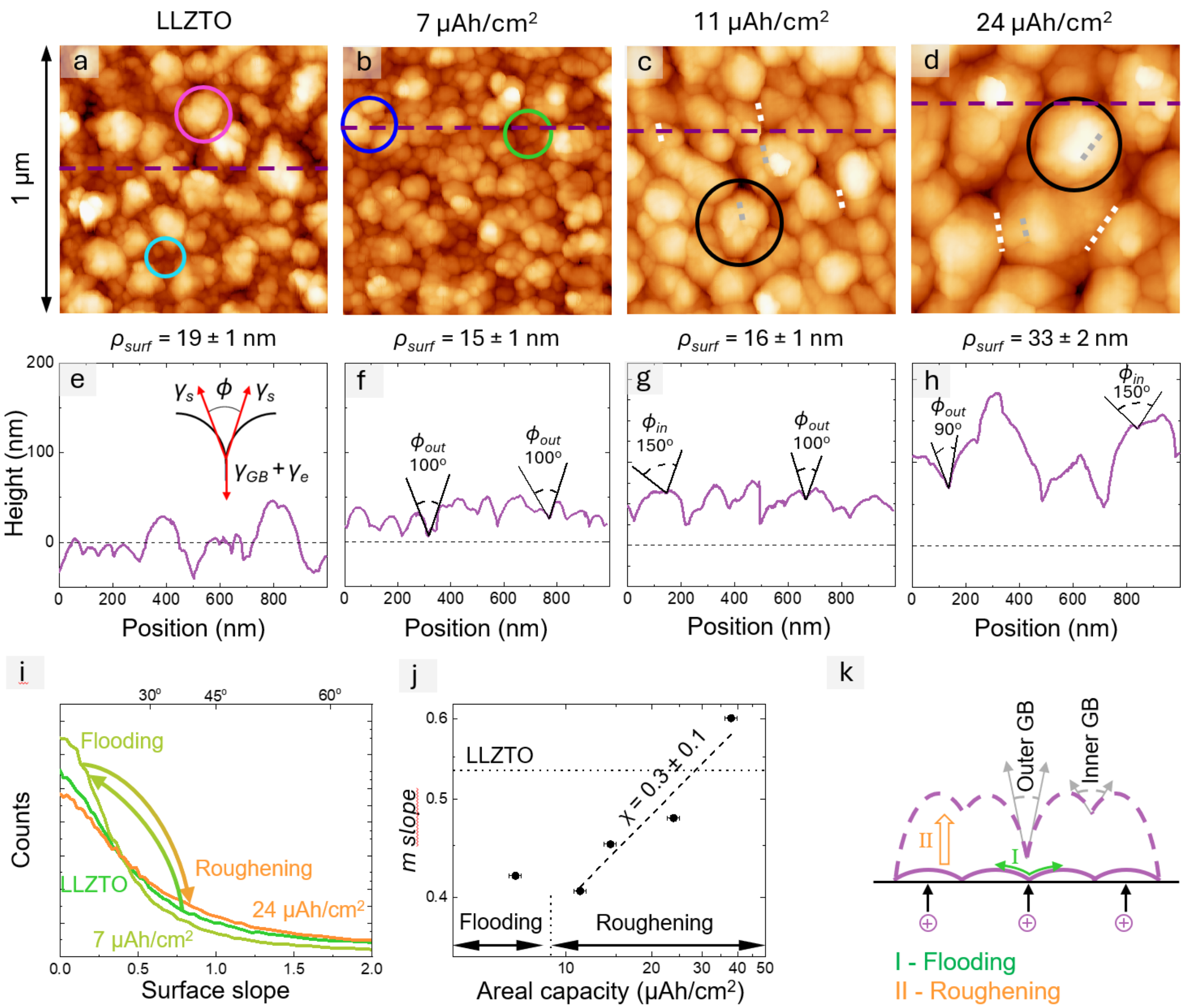

**Figure 4 | Early stages of Li plating.**

a, LLZTO surface topography showing large mounds (~150 nm) and finer background grains (~45 nm).

b–d, AFM images illustrating the transition from substrate flooding to surface roughening during Li plating.

e–h, Height profiles illustrating the evolution from conformal flooding to roughened cluster growth; $\phi_{in}$ and $\phi_{out}$ denote dihedral angles at inner and outer grain boundaries, respectively.

i,j, Slope distributions and mean surface slope distinguishing a flooding regime (decreasing roughness and slope) from the roughening regime (concurrent increase).

k, Schematic of the transition from substrate flooding to cluster coarsening and roughening, driven by surface and grain-boundary energetics.

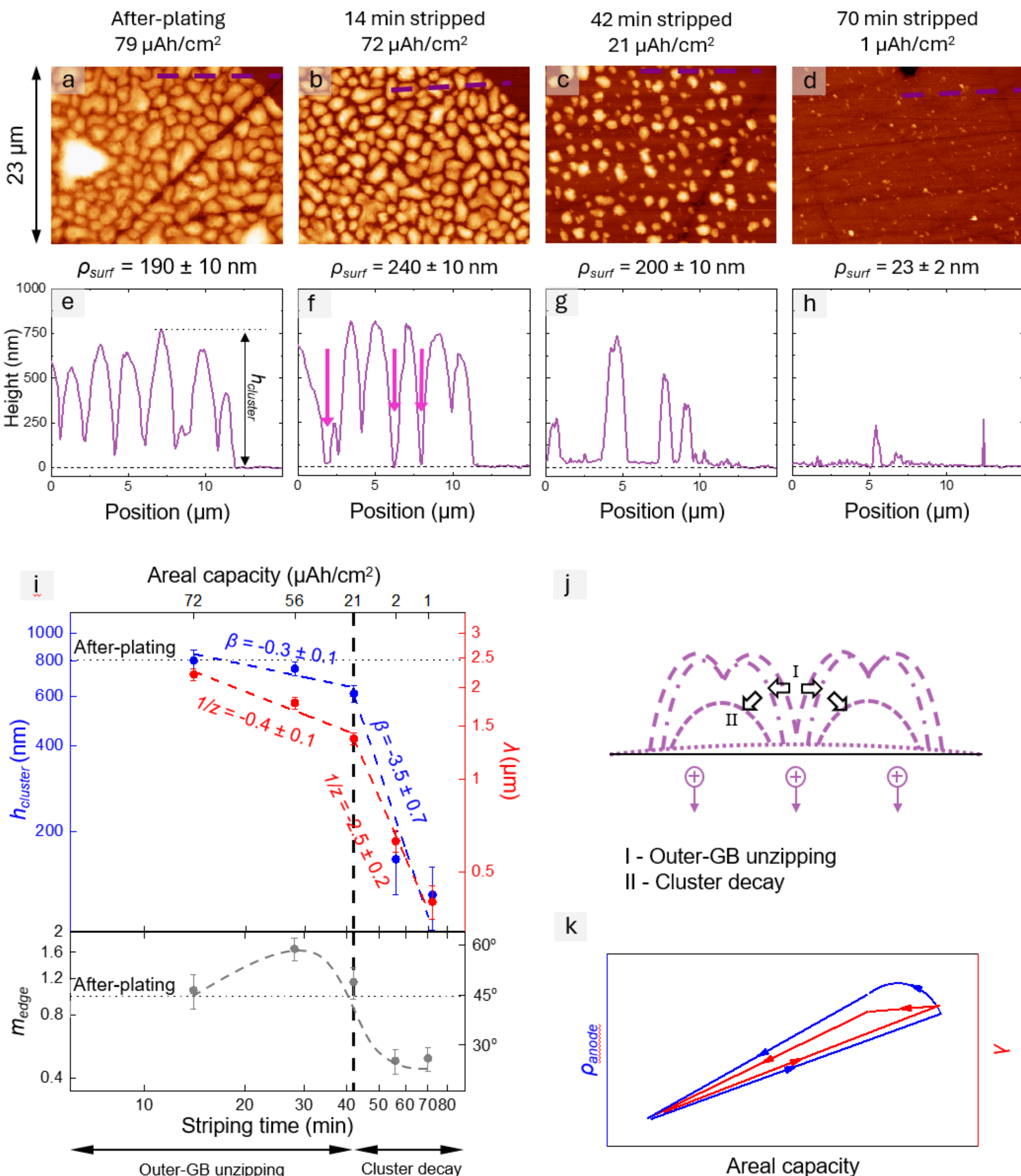

**Figure 5 | Dynamics of Na stripping.**

a, Initial Na anode morphology after plating (79 μAh cm$^{-2}$).

b–d, AFM images after 14, 42, and 70 min of UV-induced stripping, showing progressive trench deepening and cluster decay.

e–h, Corresponding height profiles.

i, Evolution of cluster height ($h_{cluster}$), lateral size ($\lambda$), and edge slope ($m_{edge}$), delineating two regimes: outer-GB unzipping and cluster decay.
j, Schematic of the stripping sequence highlighting the formation of a persistent residual layer at the electrolyte interface.
k, Schematic comparison of evolution of roughness and cluster size over a full plating–stripping cycle.
Na stripping proceeds through sequential, non-mirror processes that leave behind a residual interfacial layer, which may limit reversibility.

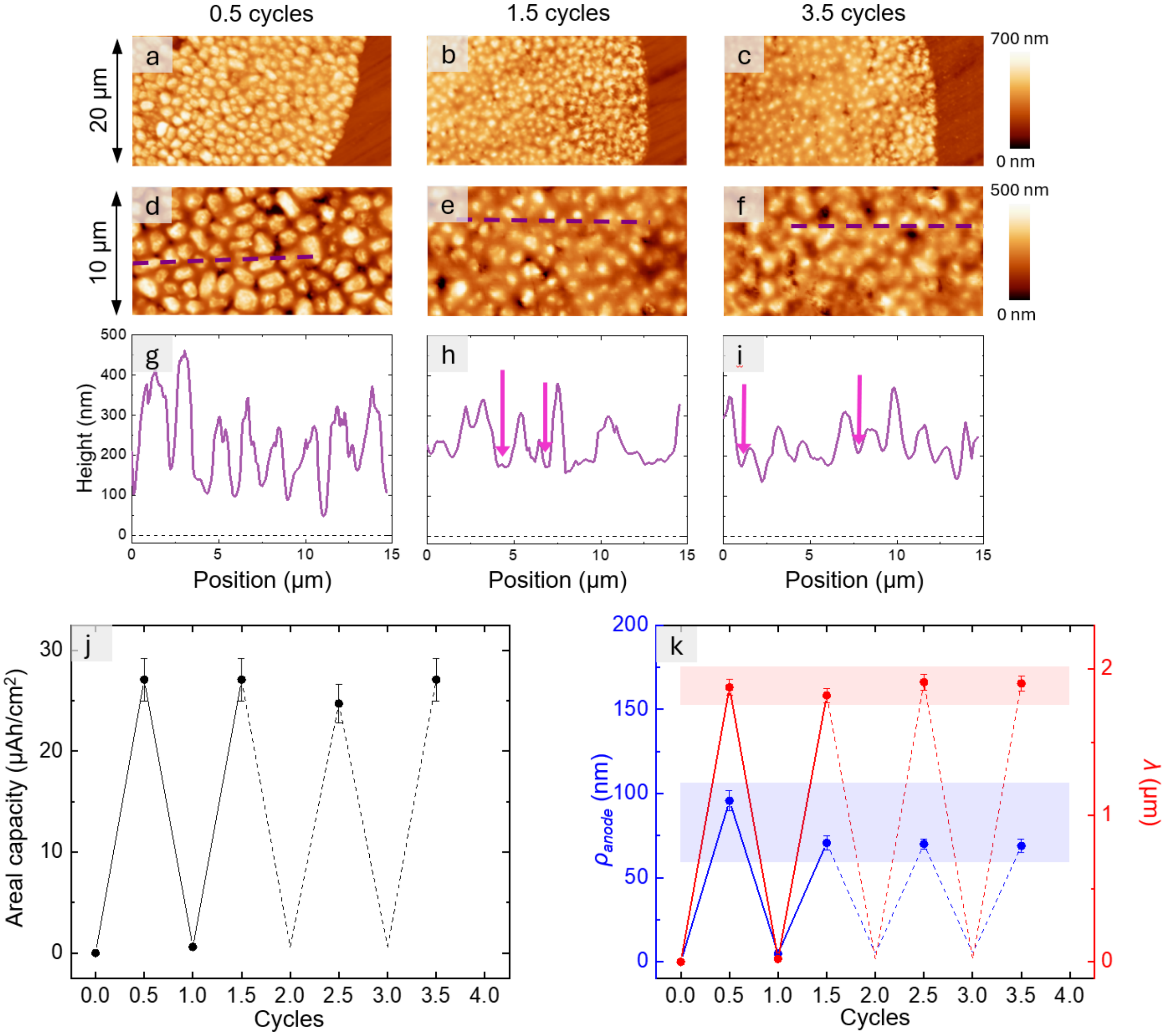

**Figure 6 | Cycling reversibility under VE-LEEM.**

a–f, AFM images of Na anode morphology after 0.5, 1.5, and 3.5 cycles, showing recovery of surface features after each plating step.

g–i, Height profiles illustrating shallower grain-boundary trenches in subsequent cycles, consistent with the presence of a residual layer.

j,k, Evolution of areal capacity, anode roughness ($\rho_{anode}$), and cluster size ($\lambda$) over four cycles, showing reproducible nucleation and slight smoothing after the first stripping event. Data points at cycle numbers 2 and 3 are extrapolated and set equal to the values after one full cycle, as the capacity and anode roughness returns to approximately zero after a complete cycle within experimental uncertainty.

VE-LEEM cycling yields largely reversible morphological evolution while indicating that a persistent, thin residual layer can subtly redirect subsequent growth without immediate capacity loss.

**Supplementary information**

# Nanoscale imaging reveals critical plating and stripping mechanisms in anode-free lithium and sodium solid-state batteries

## Overview

This Supplementary Information establishes the robustness, reproducibility, and physical consistency of the nanoscale plating and stripping mechanisms reported in the main text. All analyses are referenced to AFM-derived areal capacity, enabling direct comparison across beam conditions without requiring assignment of device-scale current densities.

Sections are organized as follows:

- **S1.** Optimization of electron-beam parameters for VE-LEEM plating
- **S2.** Quantification of areal capacity
- **S3.** Time dependence of capacity during plating and stripping
- **S4.** PEEM spectroscopy and chemical-state assignment
- **S5.** AFM analysis methodology
- **S6.** Aspect ratio and dihedral-angle analysis
- **S7.** Energetic model for grain-boundary energy estimation

## S1. Optimization of electron-beam parameters for VE-LEEM plating

To ensure that VE-LEEM induces controlled alkali-metal deposition rather than beam-induced damage or charging artifacts, we systematically optimized the incident electron energy and emission current.

### Electron kinetic energy

Lithium plating experiments were conducted on single-crystal LLZTO using electron kinetic energies between 3 and 50 eV at a fixed emission current of 1 μA for 30 min, corresponding to an areal capacity of 66 μAh cm$^{-2}$ (Fig. S1). For energies between 6 and 15 eV, well-defined and spatially stable anode regions were obtained. At 3 eV, plating efficiency was reduced, yielding smaller clusters which we attribute to a reduced effective electron flux reaching the surface due to partial surface charging that 3 eV electrons cannot robustly overcome. At 20 eV, the anode contrast was strongly suppressed, and at 50 eV pronounced instabilities and irreversible electrolyte degradation were observed.

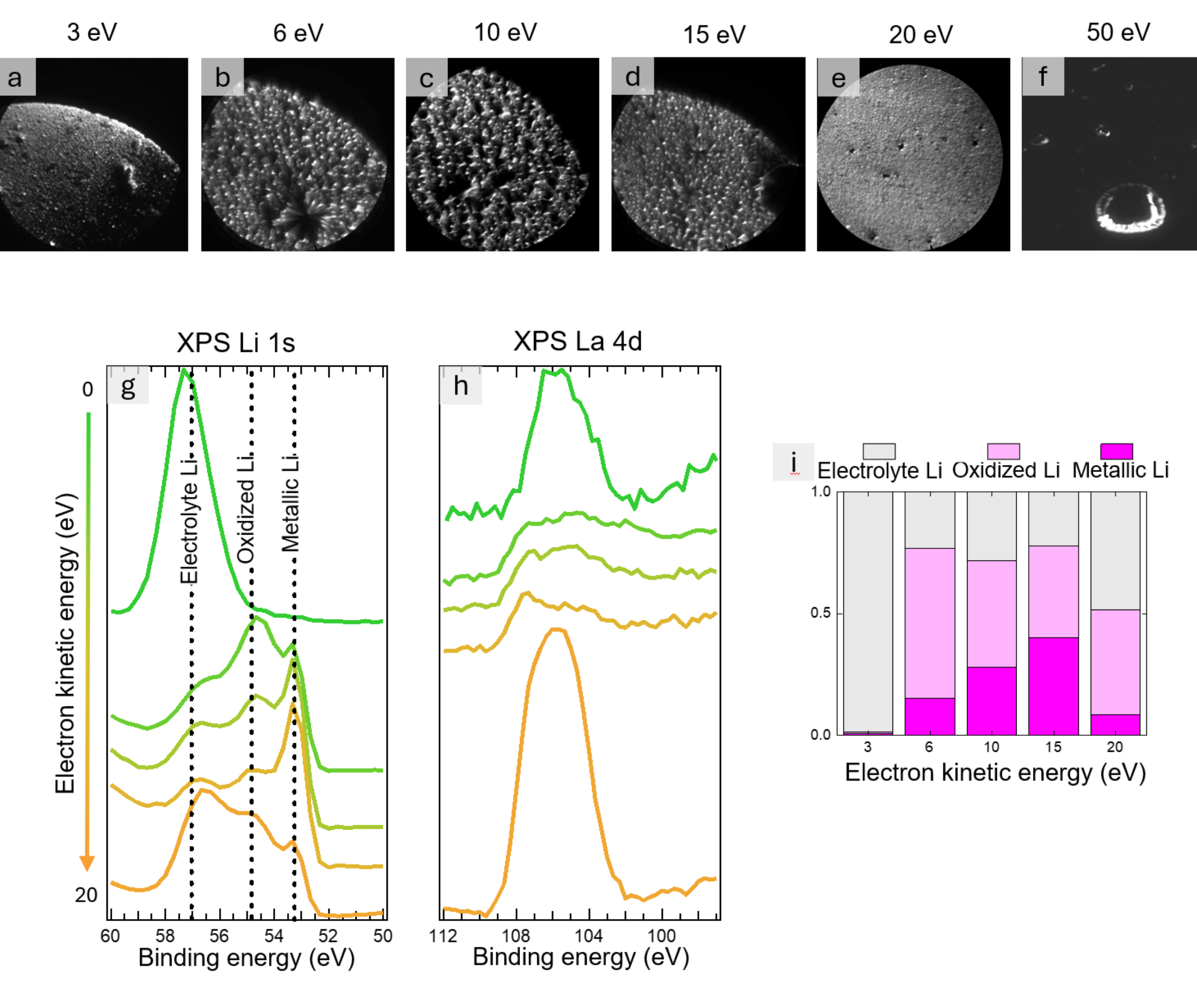

**Figure S1**

PEEM–XPS spectra acquired from the plated regions show that anodes grown between 6 and 15 eV exhibit comparable total deposited thickness (metallic Li plus oxide contributions), whereas growth at 3 and 20 eV results in significantly thinner deposits. Notably, the relative oxide contribution is slightly reduced at 15 eV, which may indicate the onset of electron-beam–induced chemical reactions, such as partial reduction of surface lithium oxide. Given the well-known propensity of metallic lithium to oxidize even under $10^{-9}$–$10^{-10}$ mbar UHV conditions, variations in the apparent oxide fraction may also reflect non-systematic effects such as small pressure fluctuations or differences in the delay between growth and spectroscopic acquisition. To minimize the likelihood of beam-induced chemical modification while ensuring homogeneous, damage-free plating, all experiments in this work were therefore performed at an electron kinetic energy of 6 eV.

**Emission current**

To evaluate the influence of electron flux, sodium plating experiments on NaGdSiO were performed at 6 eV using emission currents of 0.01, 0.1, and 1 µA (Fig. S2). Irradiation times were adjusted to deliver comparable total injected charge. At very low emission current (0.01 µA), reduced areal capacity and smaller clusters were observed, attributed to partial electronic leakage through the electrolyte during prolonged exposure. Within the validated operating window (≥ 0.1 µA), anode morphology and capacity depend primarily on total injected charge rather than emission current. Although higher emission currents may be possible in principle, the present study was limited to ≤ 1 µA, which ensured stable, damage-free and reversible operation under the conditions employed.

For comparative lithium and sodium experiments in the main text, different emission currents were selected to achieve practical deposition times under damage-free conditions; all mechanistic comparisons are made at fixed areal capacity and are therefore insensitive to this choice.

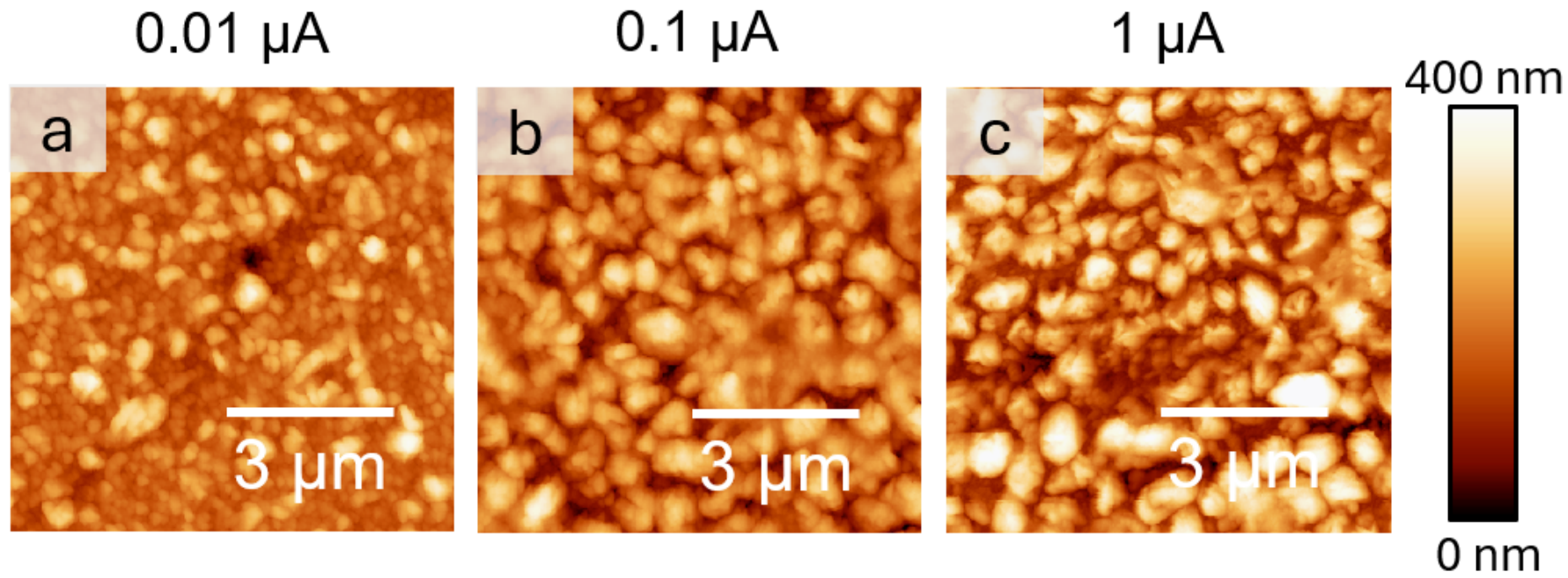

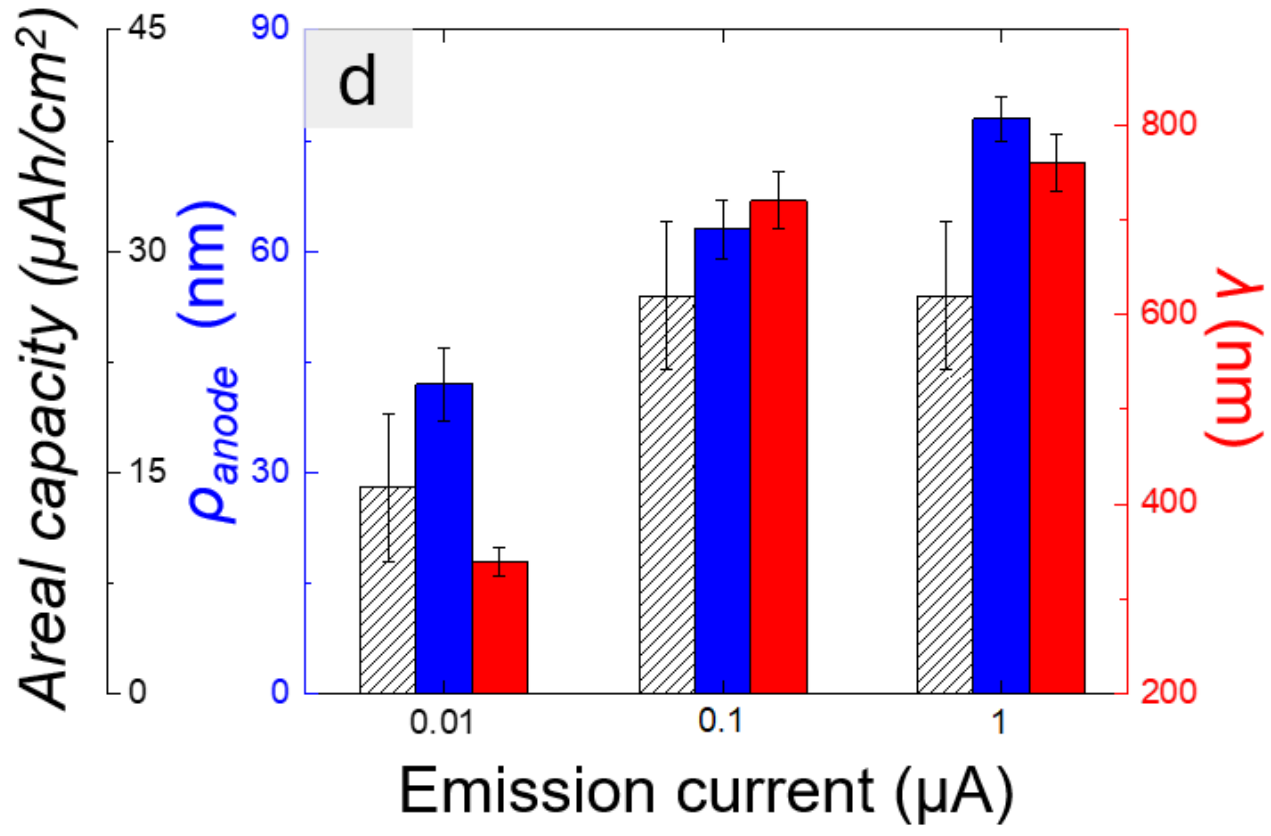

**Figure S2**

## S2. Quantification of areal capacity

The effective areal capacity $Q$ is defined as the charge per unit area required to form an alkali-metal layer of average thickness $h$:

$$Q = \rho \, h \, q,$$

where $\rho$ is the atomic density of the metal and $q$ is the ionic charge. Thickness values are extracted from AFM topography and averaged over the analyzed area. This definition yields an effective capacity suitable for scaling analysis and morphological comparison. It does not aim to provide an absolute electrochemical capacity, as effects such as nanoscale porosity or incomplete local filling are implicitly averaged in the AFM-derived thickness.

## S3. Time dependence of capacity during plating and stripping

Figure S3 shows the evolution of the sodium anode areal capacity during VE-LEEM plating and subsequent UV-induced stripping. For clarity of comparison, the capacity curves are normalized to unity for each process, enabling a direct comparison of the temporal evolution during plating and stripping despite their different absolute time scales and charge fluxes. During plating at 6 eV and 0.3 μA, the areal capacity increases approximately linearly with time, demonstrating a stable reduction rate under constant electron flux. This near-linearity is consistent with a local galvanostatic growth regime, indicating that charge transport through the growing Na layer does not impose a significant kinetic limitation under the conditions employed. Minor deviations from perfect linearity are attributed to uncertainties in the AFM-based thickness determination and morphological heterogeneity of the plated layer.

During stripping, the capacity–time dependence can also be described to first order by a linear fit when excluding the final data point associated with the persistence of the residual interfacial layer, which causes an apparent saturation of the normalized capacity at late times. In contrast to plating, however, the extracted charge flux during UV-induced stripping is not strictly constant, as the photoemission yield and local charge extraction efficiency evolve with the changing surface morphology and electronic structure of the Na anode. In particular, the progressive reduction of PEEM-UV intensity during stripping indicates a decreasing effective charge flux at later stages. Consequently, small deviations from linearity may reflect both experimental averaging effects in the AFM thickness measurement and genuine variations in the stripping kinetics as the morphology evolves.

Taken together, these results indicate that, to first order, both plating and stripping exhibit an approximately linear dependence of areal capacity on time under VE-LEEM conditions. The normalization highlights the similar gross temporal scaling of the two processes, while the fundamental asymmetry between plating and stripping arises from their distinct underlying mechanisms rather than from their overall time dependence. Specifically, plating proceeds via constant-flux, reduction-driven growth, whereas stripping is governed by morphology- and photoemission-coupled charge extraction, consistent with the non-mirror morphological pathways discussed in the main text.

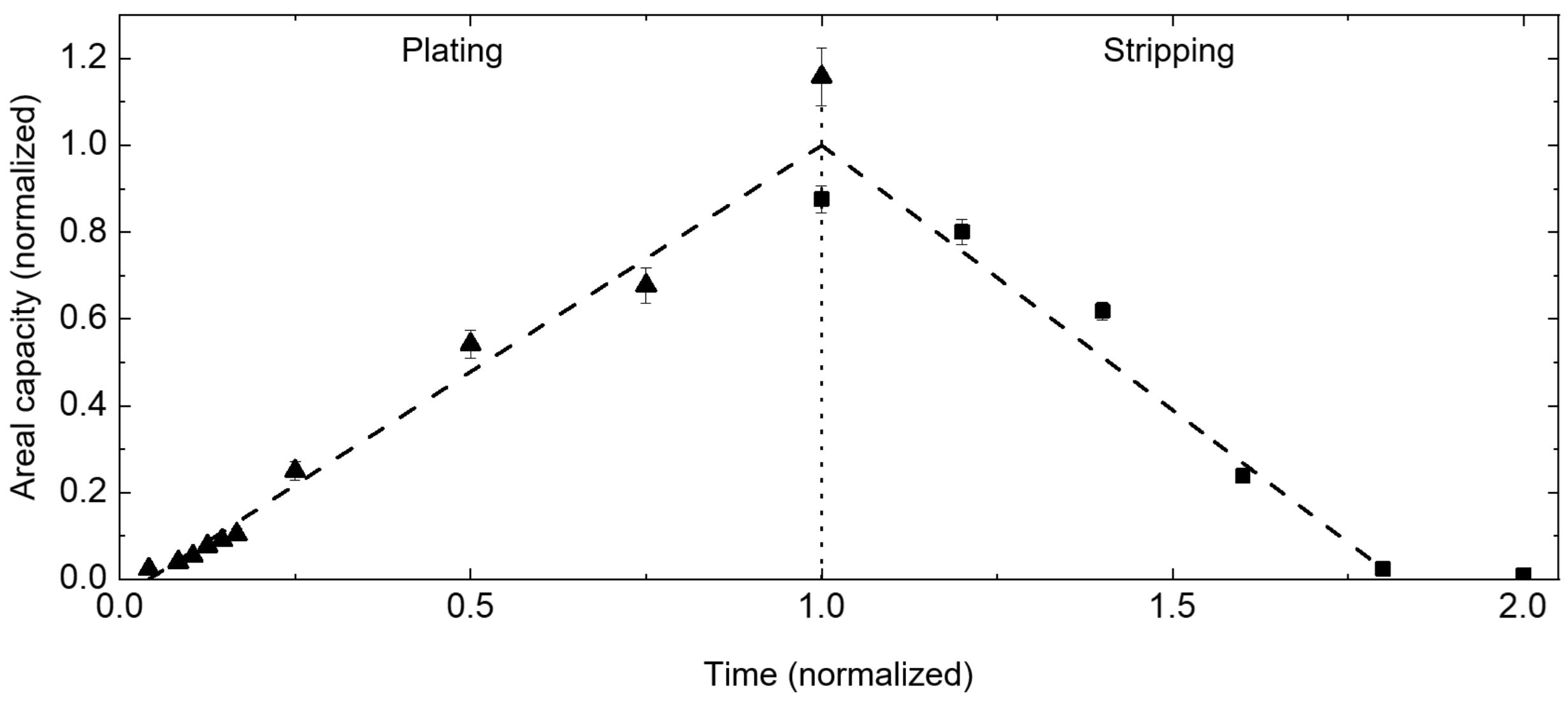

**Figure S3**

## S4. PEEM spectroscopy and chemical-state assignment

The evolution of Na K-edge XAS and Li 1s XPS spectra with increasing areal capacity confirms that plated regions are predominantly metallic (Fig. S4). For sodium, electrolyte-related features, characterized by peaks at 1074.5 eV and 1078.5 eV corresponding to $Na^+$ 1s→3p transitions, progressively decrease as a metallic, step-like absorption onset emerges at 1071.5 eV. For lithium, the electrolyte-related contribution at 57 eV diminishes, while the metallic Li 1s component at 53.5 eV increases. Intermediate components at 54.75 eV are attributed primarily to surface oxidation of alkali metals under limited UHV conditions and may additionally include contributions from interfacial species. Spectral fitting distinguishes the dominant chemical states rather than resolving subtle interphase compositions.

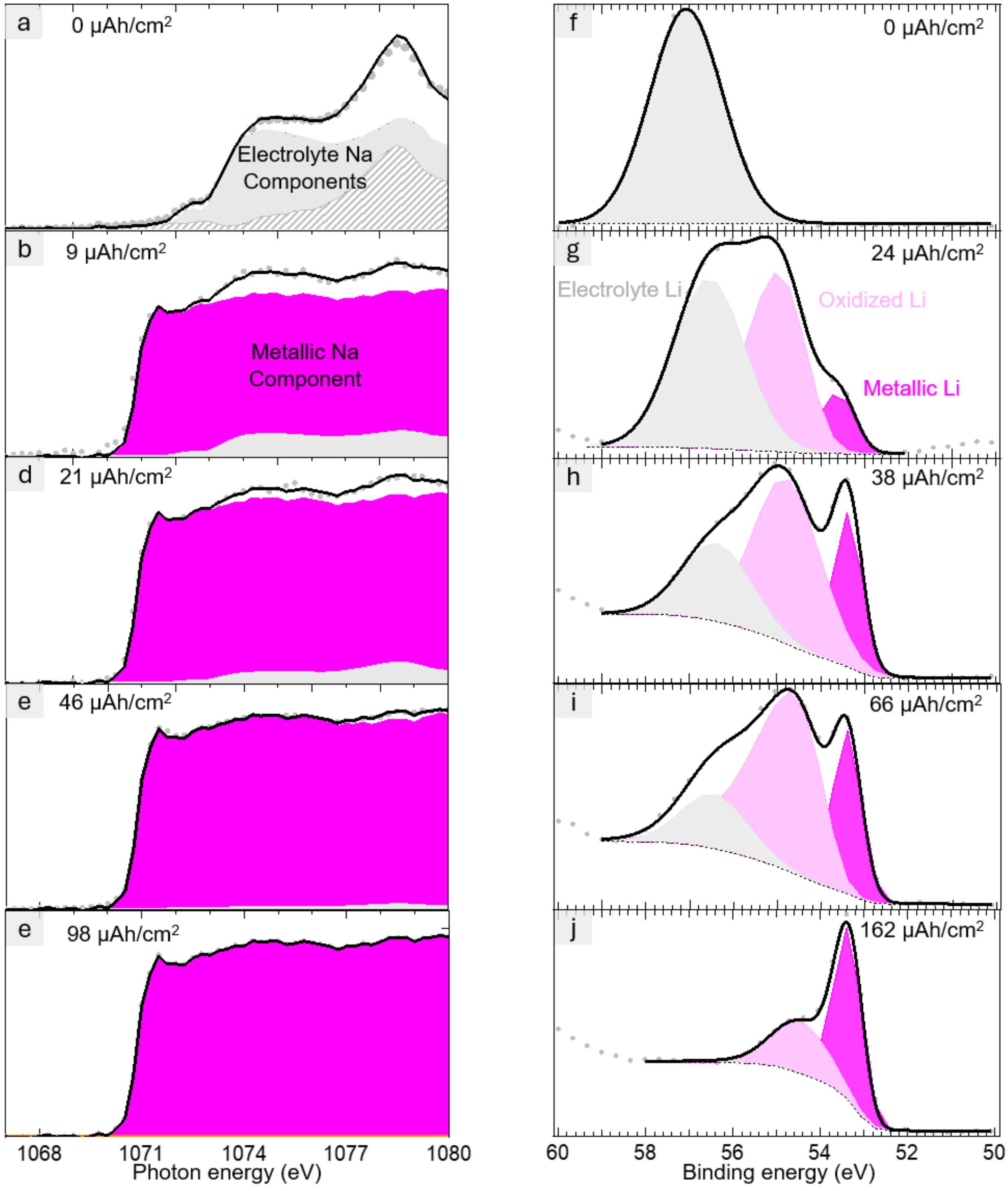

**Figure S4**

## S5. AFM analysis methodology

AFM images of pristine LLZTO reveal a bimodal topography consisting of large mounds (pink circle in Fig. S5a) and a fine-grained background (blue circle in Fig. S5a). Height-distribution analysis yields a background roughness contribution of $\rho_{electrolyte_bg}$ = 9.2 ± 0.9 nm (Fig. S5e). Panels S5f–S5h show the height-distribution analysis of the AFM topographies in Fig. S5b-S5d.

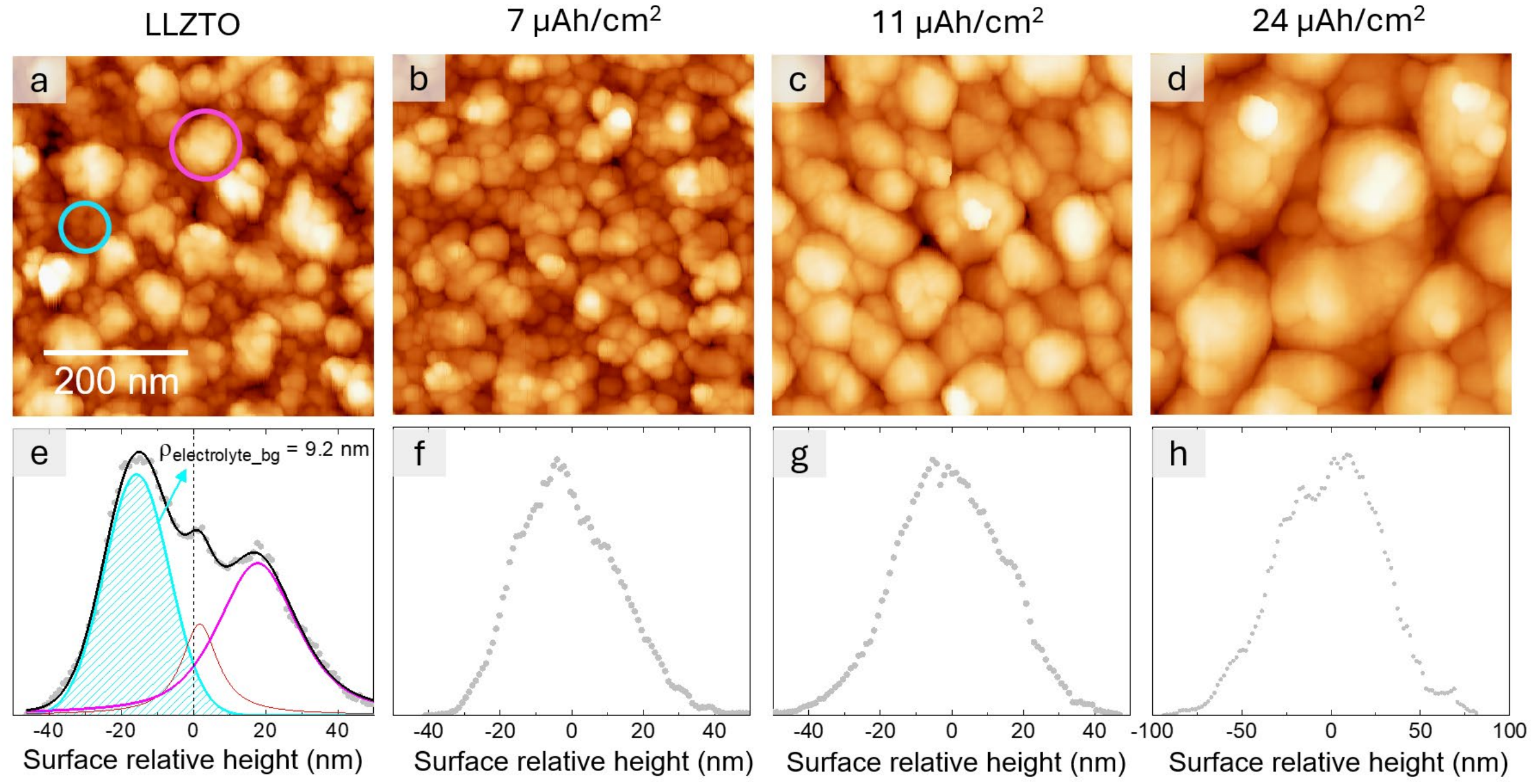

**Figure S5**

The Li anode roughness was calculated as

$$\rho_{anode} = \sqrt{\rho_{surf}^2 - \rho_{electrolyte_bg}^2}$$

The Na anode roughness was calculated as

$$\rho_{anode} = \sqrt{\rho_{surf}^2 - \rho_{electrolyte}^2}$$

where $\rho_{electrolyte}$ denotes the roughness of the pristine NaGdSiO surface measured under identical imaging conditions.

The characteristic cluster size $\lambda$ for Li and Na anodes during plating was determined by two independent methods: (i) direct cluster counting and (ii) autocorrelation analysis. For square AFM images of side length $l$, direct counting yields

$$\lambda = \frac{2l}{\sqrt{\pi N}},$$

where $N$ is the number of clusters within the imaged area. In the second approach, $\lambda$ was extracted from the first minimum of the radially averaged autocorrelation function of the AFM topography. Both methods provide consistent and comparable values of $\lambda$.

During Na anode stripping, the characteristic cluster size $\lambda$ was determined by two independent methods: (i) direct measurement from line profiles extracted from AFM images, and (ii) indirect estimation from the cluster-covered area fraction obtained using a flooding-based segmentation procedure. Both methods yielded consistent and comparable values of $\lambda$.

## S6. Aspect ratio and dihedral-angle analysis

Anode roughness $\rho_{anode}$ scales linearly with cluster height $h_{cluster}$ for both Li and Na (Fig. S6a-b), indicating a constant cluster aspect ratio $AR_{cluster} = h_{cluster}/\lambda$ throughout the growth (Fig. S6c-d). The dihedral angles of outer grain boundaries $\phi_{out}$ remain constant within experimental uncertainty (Fig. S6e-f), consistent with self-similar growth governed by surface and grain-boundary energetics.

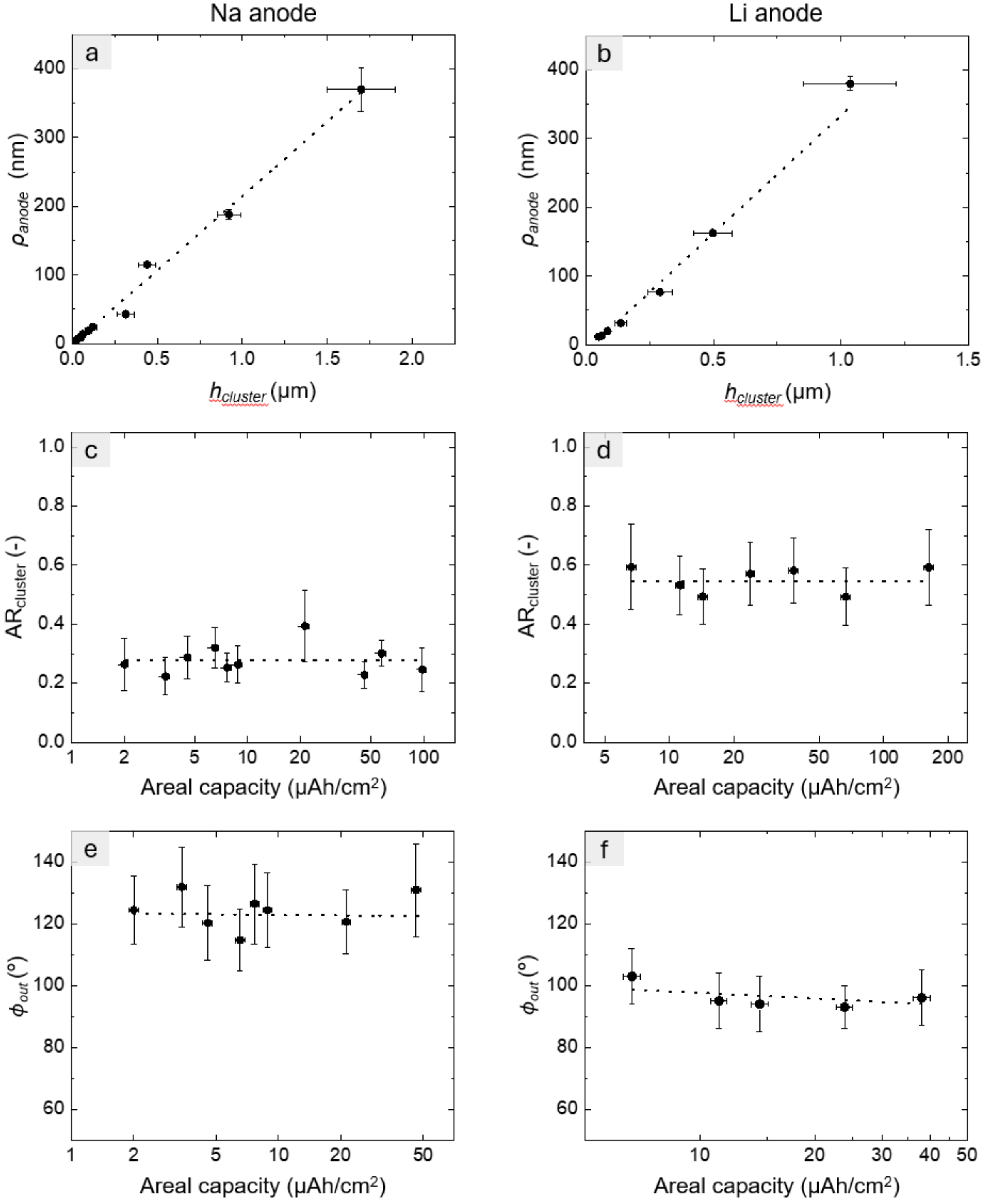

**Figure S6**

## S7. Energetic model for grain-boundary energy estimation

Surface-slope distributions during lithium plating (Fig. 4i) provide a geometric basis for estimating relative grain-boundary (GB) energies. Assuming quasi-equilibrium boundary geometries at the triple junction between two Li surfaces and a GB, the dihedral angle $\phi$ is related to the GB and surface energies through a Young-Herring force balance. Using a Li surface energy $\gamma_S^{Li}$ = 0.46 J·m$^{-2}$, the estimated GB energies are consistent with the observed transition from substrate flooding to roughening:

**(a) Invariant pivot slope (strain-free boundary).**

The slope distribution consistently pivots around a slope value $m_0$=0.48 (dihedral angle $\phi_0 = 128.7°$), whose probability remains unchanged across both growth regimes. This invariance suggests that $m_0$ corresponds to a strain-free GB configuration, for which the elastic contribution vanishes, $\gamma_e(\phi_0) \approx 0$. Under this assumption, the corresponding effective strain-free GB energy is:

$$\gamma_{GB}^0 = 2 \cdot \gamma_S^{Li} \cdot cos\left(\frac{\phi_0}{2}\right) \approx 0.86 \cdot \gamma_S^{Li}.$$

**(b) Slightly misoriented inner boundaries (CSL-like).**

For boundaries between slightly misoriented grains within clusters, the effective GB energy (including elastic contributions) is estimated as:

$$\gamma_{GB}^{CSL} + \gamma_e = 2 \cdot \gamma_S^{Li} \cdot cos\left(\frac{\phi_{in}}{2}\right) \approx 0.52 \cdot \gamma_S^{Li}.$$

Here, $\gamma_{GB}^{CSL}$ denotes an effective energy scale associated with low-Σ CSL or pseudocoherent boundaries inferred from the slope evolution.

**(c) Highly misoriented outer boundaries.**

For boundaries between highly misoriented clusters, the combined GB and elastic energy is:

$$\gamma_{GB} + \gamma_e = 2 \cdot \gamma_S^{Li} \cdot cos\left(\frac{\phi_{out}}{2}\right) \approx 1.23 \cdot \gamma_S^{Li}.$$

These estimates are intended as order-of-magnitude consistency checks rather than precise interfacial energies, given the non-equilibrium nature of electrochemical growth and the possible anisotropy of $\gamma_S^{Li}$ and GB energies.